\documentclass[superscriptaddress,secnumarabic,amssymb,amsmath,nobibnotes,aps,prd,nofootinbib,preprint]{revtex4}

\usepackage{graphicx}
\usepackage{float}
\usepackage{bm}
\usepackage{amsmath}
\usepackage{amsfonts}
\usepackage{amssymb}
\usepackage{epstopdf}
\usepackage{natbib}%
\newcommand{\bee}{\begin{equation}}
\newcommand{\eee}{\end{equation}}
\newcommand{\eaa}{\end{eqnarray}}
\newcommand{\baa}{\begin{eqnarray}}

\usepackage{color}

\begin{document}
\title{Revisiting the Immirzi parameter: Landauer's principle and alternative entropy frameworks in Loop Quantum Gravity}                          

\author{Everton M. C. Abreu}
\email{evertonabreu@ufrrj.br}
\affiliation{Departamento de F\'isica, Universidade Federal Rural do Rio de Janeiro, RJ, Brazil}
\affiliation{Applied Physics Graduate Program, Physics Institute, Federal University of Rio de Janeiro, RJ, Brazil}
\author{Jorge Ananias Neto}
\email{jorge.ananias@ufjf.br}
\affiliation{Departamento de F\'isica, Universidade Federal de Juiz de Fora, Juiz de Fora, MG, Brazil}
\author{Ronaldo Thibes}
\email{thibes@uesb.edu.br}
\affiliation{Universidade Estadual do Sudoeste da Bahia, DCEN, Itapetinga, BA, Brazil}

\begin{abstract}
This paper investigates the implications from area quantization in Loop Quantum Gravity, particularly focusing on the application of 
the Landauer principle -- a fundamental thermodynamic concept establishing a connection between information theory and 
thermodynamics. By leveraging the Landauer principle in conjunction with the Bekenstein-Hawking entropy law, we derive the usual 
value for the Immirzi parameter precisely, 
$\gamma = \ln2/(\pi \sqrt{3})$, without using the typical procedure that involves the Boltzmann-Gibbs entropy.
Furthermore, following an analogous procedure, we derive a modified expression for the Immirzi parameter aligned with 
Barrow’s entropy formulation. 
Our analysis also yields a new expression for the Immirzi parameter consistent with a corresponding modified Kaniadakis entropy 
for black hole entropy 
further illustrating, along with Barrow’s entropy, the applicability of Landauer's principle in alternative statistical contexts 
within black hole physics.
\end{abstract}

\maketitle

\section{introduction}
The quantum operator for area in Loop Quantum Gravity (LQG) has a discrete spectrum \cite{lqg}. This result plays a key role in 
the overall framework of 
LQG and suggests that geometric quantities such as area and volume, unlike in classical general relativity, are quantized at the Planck 
scale. In classical 
treatments of gravity, spacetime is considered to be smooth and continuous. However, the discrete nature of the area operator in LQG implies that 
spacetime may have an underlying granular structure at small scales, a significant departure from classical concepts.

In LQG, the quantum states of the gravitational field are described by a Hilbert space that is spanned by spin networks. Spin networks are graphs, 
with vertices and edges, where the edges are labeled by quantum numbers corresponding to the irreducible representations of the SU(2) group. 
These quantum numbers, denoted here by $j$, take half-integer or integer values $j = 0, 1/2, 1, 3/2, \dots \;$. The spin network structure 
encodes the quantum geometry of space, with the edges representing quantum excitations of the gravitational field.

The area of a given region of space in LQG is determined by the intersections of the spin network edges with a surface. In the terminology of LQG, 
these intersections are referred to as ``punctures". Each puncture contributes a quantized amount of area to the total area of the surface. 
The contribution from a puncture labeled by a spin $j$ is proportional to a function of $j$, and the total area is the sum of contributions 
from all punctures on the surface.
Mathematically, the area element associated with a spin network edge labeled by $j$ can be expressed as \cite{lqg2,lqg3,lqg4,immi,barb}
\begin{eqnarray}
\label{aj}
a(j) = 8 \pi l_p^2 \gamma \sqrt{j(j+1) } \,,  
\end{eqnarray}
where $\gamma$ denotes the Barbero-Immirzi parameter \cite{immi,barb} and $l_p^2$ the Planck length.  In principle, the Barbero-Immirzi 
parameter $\gamma$, also known as Immirzi parameter, is a dimensionless real number characterizing an open fundamental quantity within 
the theory.  Eq. (\ref{aj}) shows that the 
area contribution increases with the quantum number $j$ and, for each puncture, the area is quantized in discrete units.
A specific expression can be derived for the Immirzi parameter in the context of Boltzmann-Gibbs (BG) statistics, as we show next 
following \cite{dreyer}.

The number of configurations (microstates) on a punctured surface is given by
\begin{eqnarray}
\label{immi}
 W = \prod_{n=1}^N \left(2 j_n + 1\right) \,,
\end{eqnarray}
where the factor $(2 j_n + 1) $ in Eq. (\ref{immi}) is the multiplicity of the state $j_n$.
It can be demonstrated that the primary configurations that contribute to Eq. (\ref{immi}) are those in which the spin assumes its 
minimum possible 
value. Denoting $j_{min}$ as this lowest spin value, we then find from Eq. (\ref{immi}) that
\begin{eqnarray}
\label{mini}
 W = \left( 2 j_{min} + 1\right)^N \,.
\end{eqnarray}

From Eq. (\ref{aj}), the number of punctures on a surface with total area $A$ can be determined as
\begin{eqnarray}
\label{n}
 N = \frac{A}{\Delta A} = \frac{A}{8 \pi l_p^2 \gamma \sqrt{j_{min} (j_{min}+1)} }  \,,
\end{eqnarray}
where we have naturally assumed that
\begin{eqnarray}
\label{da}
\Delta A = a(j_{min}) \,,
\end{eqnarray}
according to Eq. (\ref{aj}).

In Boltzmann-Gibbs (BG) statistics, the entropy is expressed as $S=k_B \ln W$ which, from Eq. (\ref{mini}), leads to 
\begin{eqnarray}
\label{sbg}
 S = k_B N \ln (2 j_{min} +1)  \,.
\end{eqnarray}
Finally,
by equating the well-known Bekenstein-Hawking entropy area law $S = \displaystyle  k_B \frac{A}{4l_p^2}$ 
to Eq. (\ref{sbg}) and using Eq. (\ref{n}), with $j_{min} = \frac{1}{2}$, we derive the forementioned Immirzi parameter relation 
within the BG statistics context as
\begin{eqnarray}
\label{immip}
 \gamma = \frac{\ln 2}{\pi \sqrt{3}}  \,.
\end{eqnarray}

\section{Landauer principle}
The Landauer principle \cite{landauer,bri,land2}, introduced by Rolf Landauer in 1961, is a fundamental concept that integrates ideas 
from information theory and thermodynamics. It states that erasing a single bit of information requires a minimum energy 
expenditure, mathematically described by the inequality
\begin{eqnarray}
 \label{land}
 \Delta E \geq k_B T \ln 2 \,,
\end{eqnarray}
where $k_B$ is the Boltzmann constant and $T$ is the absolute temperature.
The $\ln 2$ term in Eq. (\ref{land}) arises from the fact that the entropy of an information system, which measures its disorder, 
is related to the number of possible microstates, given by $\Omega = 2^N$, where $N$ represents the number of bits. The energy 
described in Eq. (\ref{land}) is released as heat into the environment.
Landauer’s principle showcases the physical aspect of information and its fundamental link to thermodynamics. Any process involving 
information manipulation is subject to thermodynamic laws.  Thus, erasing information produces heat, in line with 
the second law of thermodynamics, which states that the entropy of an isolated system cannot decrease.

Despite occasional challenges to its validity, Landauer’s principle has remained robust under extensive scrutiny. 
Different experiments have consistently confirmed the energy cost of erasing information, closely matching the corresponding 
theoretical predictions \cite{bap,jgb,yxr}. Furthermore, rigorous mathematical proofs have strengthened 
the principle \cite{rw,ebor,ebor2}, solidifying 
its role as a fundamental concept in the thermodynamics of computation.
The recent article \cite{cl} presents a compelling application of Landauer's principle to black hole physics, specifically in relation to Hawking 
evaporation. The authors of \cite{cl} observe that Hawking evaporation precisely satisfies Landauer's principle, with the information 
lost by the black hole occurring in a thermodynamically optimal manner. Notably, in the beginning of the year, Abreu has established a 
connection between the Hawking temperature and Landauer's principle \cite{abreu}. Building on that foundation, further intriguing 
works have extended 
the application of Landauer’s principle to the cosmological horizon \cite{otr} and explored its relevance to black hole area quantization \cite{bgs}.

In this context, Landauer's principle can be derived from the Bekenstein-Hawking area entropy law for black holes.  
For simplicity, we adopt a unit 
system in which $\hbar = c = 1$. Accordingly, the entropy of a black hole can be expressed in terms of its mass as
\begin{eqnarray} 
\label{bhe} 
S = 4 \pi k_B G M^2 \,, 
\end{eqnarray} 
where $G$ denotes Newton's gravitational constant. 
Using Eq. (\ref{bhe}), the temperature of the black hole can be obtained from
\begin{eqnarray} 
\label{bht} 
\frac{dS}{dM} = \frac{1}{T} = 8 \pi k_B G M \,,
\end{eqnarray}
while the entropy variation in terms of the mass loss is given by
\begin{eqnarray} 
\label{dm} 
\Delta S = 8 \pi k_B G M \Delta M \,. 
\end{eqnarray} 
Now, consider a scenario in which the black hole loses sufficient mass to reduce its information content by one bit. In that case, 
the change in entropy, as described by Landauer's principle, is 
\begin{eqnarray} 
\label{dsb} \Delta S = k_B \ln 2
\end{eqnarray}
and hence, Eq. (\ref{dm}) leads to
\begin{eqnarray} 
\label{ldm} 
\Delta M = \frac{\ln 2}{8 \pi G} \,\frac{1}{M} \,,
\end{eqnarray}
which, using Eq. (\ref{bht}), can be recast as
\begin{eqnarray} 
\label{lds} 
\Delta M = k_B T \ln 2 \,. 
\end{eqnarray} 
Eq. (\ref{lds}) represents Landauer's principle when it is saturated. Therefore, we conclude that the information lost by a 
black hole during its evaporation occurs with maximum thermodynamic efficiency.

Having derived Landauer's principle from black hole evaporation, we now turn to the Bekenstein-Hawking entropy law, 
written in terms of the black hole horizon area

\begin{eqnarray} 
\label{bhsf} 
S = k_B \, \frac{A}{4G} \,. 
\end{eqnarray} 
The variation of Eq. (\ref{bhsf}) is
\begin{eqnarray} 
\label{vbh} \Delta S = \frac{k_B}{4G} \, \Delta A \,. 
\end{eqnarray} 
Substituting Eq. (\ref{dsb}) into Eq. (\ref{vbh}), we obtain
\begin{eqnarray} 
\label{alp} 
\Delta A = 4 l_p^2 \ln 2 \,, 
\end{eqnarray} 
where we have used the fact that $G = l_p^2$ in our unit system. Next, to determine the Immirzi parameter using 
Landauer's principle, we revisit Eq. (\ref{da}). We begin by calculating $\Delta A$ for $j = 1/2$, the minimum permissible 
value of $j$. Using Eq. (\ref{aj}), we find
\begin{eqnarray} 
\label{dalqg}
\Delta A = 4 \pi l_p^2 \gamma \sqrt{3} \,. 
\end{eqnarray} 
Equating Eq. (\ref{dalqg}) with Eq. (\ref{alp}), and writing $\gamma_{Land}$ for $\gamma$,
we obtain the following value for the Immirzi parameter, as predicted by Landauer's principle
\begin{eqnarray} 
\label{gammaland} 
\gamma_{Land} = \frac{\ln 2}{\pi \sqrt{3}} \,. 
\end{eqnarray} 
From Eq. (\ref{gammaland}), 
we can observe that the Immirzi parameter $\gamma_{Land}$, derived using Landauer's principle, precisely matches the value 
obtained from the application of BG statistics to LQG, as shown in Eq. (\ref{immip}).
This result is highly meaningful and has significant implications.
It demonstrates a clear consistency between two approaches: Landauer's principle 
and BG statistics, both leading to the same value for the Immirzi parameter in LQG theory. 
Here, it is important to mention that the matching value obtained from the application of the BG statistic and the Landauer 
principle likely arises because the number of configurations (microstates) on a punctured surface, as expressed in Eq. (\ref{mini})
for $j_{min}=\frac{1}{2}$, 
is the same as the number of possible microstates, given by $\Omega = 2^N$, which is used in the derivation of the Landauer principle 
in the context of Hawking evaporation.
Thus, the precise agreement between $\gamma_{Land}$
and the value obtained from BG statistics in Eq. (\ref{immip}) suggests that Landauer's principle may also provide thermodynamic 
support for the Immirzi parameter.

\section{Barrow entropy}
In a seminal recent contribution \cite{barrow}, Barrow investigated the possibility that quantum gravitational effects might 
lead to a complex fractal structure on the surface of a black hole.  Such structure alters the actual horizon area, resulting 
in a new relation for black hole entropy, namely,
\begin{eqnarray}
\label{sbarrow}
 S_B = k_B \left( \frac{A}{ 4 l_p^2} \right)^{1+\frac{\Delta}{2}}    \,,
\end{eqnarray}
where $A$ is the usual horizon area and $\Delta$ stands for Barrow's fractal exponent \cite{barrow}.
It is important to emphasize that this extended entropy differs from the standard quantum corrected entropy, which includes 
logarithmic terms \cite{sari,sari2,aab,kj}, even though it resembles a form of Tsallis' nonextensive entropy 
\cite{tsallis,ww,tsallis2}.
The non-trivial exponent in Eq. (\ref{sbarrow}) encodes a gravitational perturbation around classical solutions due to quantum effects.
In principle $\Delta$ may vary from $0$, corresponding to the simplest model resulting in the well-known Bekenstein-Hawking entropy, 
to $\Delta = 1$, the referred maximal deformation  \cite{Abreu:2020dyu,bimmi}.

To calculate the Immirzi parameter within the framework of Barrow's entropy, let us first consider the variation in 
entropy (\ref{sbarrow})
\begin{eqnarray}
\label{dsba} 
\Delta S = k_B \, \left( 1+\frac{\Delta}{2} \right) \left( \frac{A}{4l_p^2} \right)^{\frac{\Delta}{2}}
\frac{\Delta A}{4l_p^2} \,.
\end{eqnarray} 
Using the Landauer principle, as expressed in Eq. (\ref{dsb}), which states simply that $\Delta S = k_B \ln 2$, the variation of the 
area, $\Delta A$, calculated from Eq. (\ref{dsba}) can be written as 
\begin{eqnarray} 
\label{dab} 
\Delta A = 4 l_p^2 \, \ln 2 \;  \frac{1}{1+\frac{\Delta}{2}} \left(\frac{A}{4l_p^2}\right)^{- \frac{\Delta}{2}} \,. 
\end{eqnarray} 
By equating Eq. (\ref{dab}) with (\ref{dalqg}), now writing $\gamma_{B}$ for $\gamma$, we can obtain the Immirzi parameter 
defined in the framework of Barrow entropy as
\begin{eqnarray} 
\label{gib} 
\gamma_B = \frac{\ln 2}{\pi \; \sqrt{3}} \; \frac{1}{1+\frac{\Delta}{2}} \left(\frac{4l_p^2}{A}\right)^{ \frac{\Delta}{2}}\,. 
\end{eqnarray} 
From Eq. (\ref{gib}), we can observe that in the limit $\Delta \rightarrow 0$, we recover the usual Immirzi parameter 
$\gamma = \ln 2 /\pi \sqrt{3} \,.$
\begin{figure}[H]
	\centering
	\includegraphics[height=7 cm,width=10 cm]{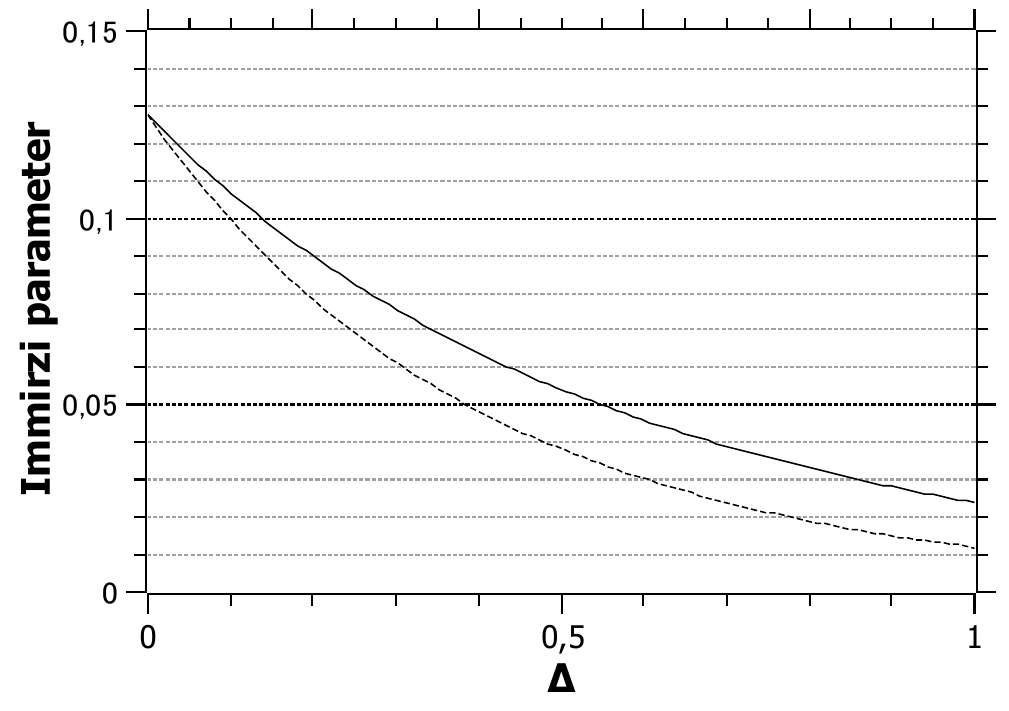}
	\caption{Values of the Immirzi parameter $\gamma_B$, Eq. (\ref{gib}), in the context of Barrow entropy, as a function of $\Delta$ for 
    areas $A=16\pi l_p^2$ (line) 
 and $A=64 \pi l_p^2$ (dash).}
	\label{barrowl}
\end{figure}
In Fig. 1, we have plotted the Immirzi parameter $\gamma_B$, Eq. (\ref{gib}), as a function of $\Delta$. The $\Delta$ parameter 
varies within the physical interval of $0 \leq \Delta \leq 1$. For this, we choose two different values for the area $A$, namely 
$A = 16 \pi l_p^2$ (line) and $A = 64 \pi l_p^2$ (dash). We can observe that, as $\Delta$ increases, the Immirzi parameter decreases 
for both curves. 

The behavior of the Immirzi parameter shown in Fig.~1 can be understood as follows. The Barrow entropy, which exceeds or equals the Bekenstein--Hawking 
entropy due to its modified mathematical structure, implies a larger number of effective microstates associated with a given horizon area. Since the 
number of microstates is given by \( 2^{\frac{A}{\Delta A}} \), where \( \Delta A \) is defined in Eq.~(\ref{da}) and satisfies \( \Delta A \propto \gamma \), 
an increase in the number of microstates requires a smaller \( \Delta A \). Consequently, the Immirzi parameter \( \gamma \) must decrease as the Barrow 
deformation parameter \( \Delta \) increases.

\section{Modified Kaniadakis entropy}

The Kaniadakis statistics, also known as $\kappa$-statistics, provides a non-extensive generalization of the 
classical Boltzmann-Gibbs (BG) framework, introducing a deformation parameter $\kappa$ that modifies the original 
entropy formulation \cite{k1,k2,k3,k4}. The $\kappa$-entropy 
is given by

\begin{eqnarray} 
S_\kappa = - k_B \sum_{i=1}^W \frac{p_i^{1+\kappa} - p_i^{1-\kappa}}{2 \kappa} \,, 
\end{eqnarray}
where $k_B$ is the Boltzmann constant, $p_i$ represents the probability of each microstate, $\kappa$ is a real parameter, and 
$W$ denotes the total number of microstates. When $\kappa \to 0$, this expression reverts to the standard BG entropy. 
The $\kappa$-entropy retains core entropy properties, except additivity, satisfying instead a pseudo-additivity condition.

The $\kappa$-entropy is especially significant for its interpretation as a relativistic extension of BG entropy and has been 
successfully applied in studies of cosmic rays, gravitational systems, and cosmic phenomena \cite{ks,kqs,aabn,aamp,lbs,ggl,as}.
In the case of a microcanonical ensemble, where each state has equal probability, the Kaniadakis entropy reduces to
\begin{eqnarray} 
S_\kappa = k_B \frac{W^\kappa - W^{-\kappa}}{2 \kappa}\,. 
\end{eqnarray}
This expression approximates the standard BG entropy formula $ S = k_B \ln W $ in the limit $ \kappa \to 0 $.

As proposed in earlier works \cite{aa}, this form can represent the black hole entropy $S_{BH}$ through

\begin{eqnarray} 
k_B \frac{W^\kappa - W^{-\kappa}}{2 \kappa} = S_{BH} \,. 
\end{eqnarray}
From this relation, we derive
\begin{eqnarray}
W = \left(\kappa \frac{S_{BH}}{k_B} + \sqrt{1+\kappa^2 \frac{S_{BH}^2}{k_B^2}}\right)^{\frac{1}{\kappa}},
\end{eqnarray}

which leads to a modified Kaniadakis entropy, given by
\begin{eqnarray}
\label{mod}
S_\kappa^* = \frac{k_B}{\kappa} \ln\left(\kappa \frac{S_{BH}}{k_B} + \sqrt{1 + \kappa^2 \frac{S_{BH}^2}{k_B^2}}\right)\,, 
\end{eqnarray}
where $S_{BH} = k_B A / {4l_p^2}$. Note that, when $\kappa = 0$, $S_\kappa^*$ reduces to $S_{BH}$.

It is important to emphasize that the Hawking temperature derived from the modified entropy in Eq.~(28) indicates that Schwarzschild 
black holes exhibit a positive heat capacity. This means that, beyond a certain critical mass, the black hole mass increases as the 
temperature rises. Such behavior contrasts with the standard case, in which the black hole mass decreases as the temperature increases. 
Moreover, by employing the modified Kaniadakis entropy, one can show that the mass loss due to radiation emission is equal to or greater 
than that predicted by the Bekenstein--Hawking entropy. This result points to a shorter evaporation time within this framework, which 
may be consistent with the non-observation of primordial black holes. Given that black hole evaporation is inherently a quantum process, 
the modified Kaniadakis entropy may serve as an effective framework for incorporating quantum corrections into black hole thermodynamics. 
For further details, see~\cite{aa,aaacnt}.

In this context, it is important to note that R\'enyi entropy has become a standard tool in quantum information theory, serving as 
a generalized measure of entanglement. It has been extensively applied to characterize entanglement spectra across event horizons 
(see, for example, \cite{calabrese,nishi}). Moreover, a modified version of the R\'enyi entropy has been used as an alternative framework 
to encode deviations from the standard entropy--area relation, with applications in black-hole thermodynamics \cite{ci,ci2,phl}. 
However, the proposal of the modified Kaniadakis entropy, often regarded as a relativistic counterpart of the modified R\'enyi entropy, 
is relatively recent~\cite{aa}. To the best of our knowledge, its only application so far in a quantum context has been in the description 
of black-hole evaporation~\cite{aaacnt}. Nevertheless, because this application alone may not be sufficient to establish a quantum nature for 
the modified Kaniadakis entropy, it is more appropriate to interpret it as an effective thermodynamic framework. In this sense, it describes 
quantum aspects of black-hole thermodynamics, as discussed in the previous paragraph. Rather than providing a possible underlying quantum theory 
of the horizon, the modified Kaniadakis entropy serves as a parametrization of deviations in the entropy--area relation that, when inserted in the 
Landauer framework, lead to the measurable consequences presented in Eqs.~(29)--(31). At present, we do not attribute quantum-entanglement status 
to $S_\kappa^\ast$, reserving such identifications to R\'enyi-based constructions. Our use of Kaniadakis entropy is phenomenological. Whether the 
modified Kaniadakis entropy could eventually account for the quantum entanglement entropy of black-hole systems remains an open question for 
future investigation.

To determine the Immirzi parameter within the framework of Modified Kaniadakis entropy, we begin by considering the variation in 
entropy (\ref{mod}), given by
\begin{eqnarray} 
\label{dsmke} 
\Delta S = \frac{\frac{k_B}{4l_p^2} \, \Delta A}{\sqrt{ 1 + \kappa^2 \frac{S^2_{BH}}{k_B^2} }} \,.
\end{eqnarray}
Applying the Landauer principle, which states that $\Delta S = k_B \ln 2$, the variation $\Delta A$
calculated from Eq. (\ref{dsmke}) can be expressed as
\begin{eqnarray} 
\label{damke} 
\Delta A = 4 l_p^2 \ln 2 \, \sqrt{ 1 + \kappa^2 \left(\frac{A}{4l_p^2}\right)^2} \,.
\end{eqnarray}
Hence, by equating relations (\ref{damke}) and (\ref{dalqg}), we can derive the Immirzi parameter within the framework of modified Kaniadakis 
entropy as
\begin{eqnarray} 
\label{imike} 
\gamma_{MKE} = \frac{\ln 2}{\pi \sqrt{3}} \, \sqrt{ 1 + \kappa^2 \left(\frac{A}{4l_p^2}\right)^2} \,.
\end{eqnarray}
From Eq. (\ref{imike}), we see that in the limit $\kappa \rightarrow 0$, the standard Immirzi parameter is recovered 
as $\gamma = \frac{\ln 2}{\pi \sqrt{3}}\,.$
\begin{figure}[H]
	\centering
	\includegraphics[height=7 cm,width=10 cm]{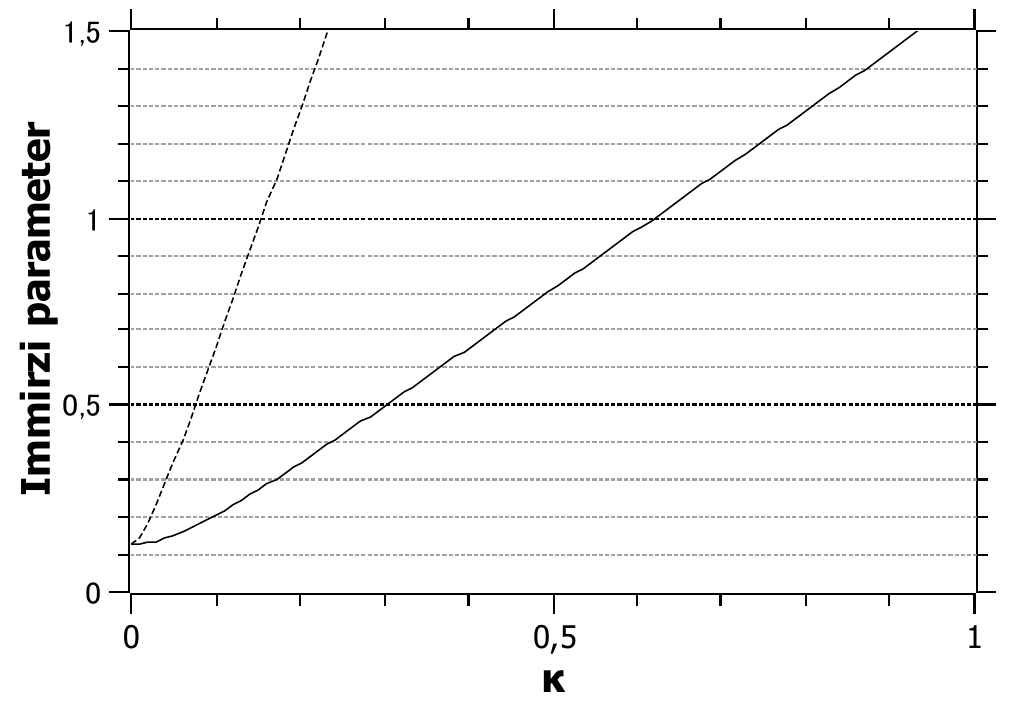}
	\caption{Values of the Immirzi parameter $\gamma_{MKE}$, Eq. (\ref{imike}), in the framework of a modified Kaniadakis entropy, as 
    a function of $\kappa$ for areas $A=16\pi l_p^2$ (line) 
 and $A=64 \pi l_p^2$ (dash).}
	\label{kaniadakisl}
\end{figure}
In Fig. 2, we present the Immirzi parameter $\gamma_{MKE}$, Eq. (\ref{imike}), plotted as a function of the variable $\kappa$, which varies 
within the physical interval $0 \leq \kappa \leq 1$. For this analysis, we consider two values for the area $A$: specifically, 
$A = 16 \pi l_p^2$ (line) and $A = 64 \pi l_p^2$(dash). It is evident that as $\kappa$ increases, the Immirzi parameter also grows in both 
curves.

The behavior of the Immirzi parameter shown in Fig.~2 can be understood as follows. The Kaniadakis entropy, which is less than or equal to 
the Bekenstein--Hawking entropy due to its modified mathematical structure, implies a smaller number of effective microstates associated 
with a given horizon area. Since the number of microstates is given by \( 2^{\frac{A}{\Delta A}} \), where \( \Delta A \) is defined in 
Eq.~(\ref{da}) and satisfies \( \Delta A \propto \gamma \), this reduction in microstates requires a larger \( \Delta A \). As a result, 
the Immirzi parameter \( \gamma \) must increase as the Kaniadakis deformation parameter \( \kappa \) increases, in contrast to the behavior 
observed in Fig.~1.

It is important to note the dependence of the Immirzi parameter on the black-hole event horizon area, as obtained in Eqs.~(\ref{gib}) and (\ref{imike}). 
Corrections to black-hole entropy arising from both the Barrow and the modified Kaniadakis entropies lead to an area dependence of the Immirzi parameter 
when the Landauer principle is applied. This emphasizes that the specific entropy framework adopted to describe the horizon plays a central role in 
determining the behavior of the Immirzi parameter. Nevertheless, allowing the Immirzi parameter to vary with the area introduces inconsistencies within 
the canonical structure of LQG and in the definition of geometric operators such as the area operator. Therefore, both the Barrow entropy and the modified 
Kaniadakis entropy should be regarded as effective entropies, providing model-dependent descriptions of quantum-related corrections to black-hole 
thermodynamics. Accordingly, the resulting $\gamma(A)$ can be interpreted as an effective quantity, while the consistency of LQG is preserved, 
with the Immirzi parameter remaining constant and the geometric operators remaining well defined.

\section{conclusions}
In this paper, we examined the effect of the Landauer principle within the framework of LQG, focusing particularly 
on the Immirzi parameter with the lowest spin possible which is $j_{min}=1/2$. Generally, the Immirzi parameter is arbitrary, 
but we propose a procedure to determine it by employing the Landauer principle. Next, from an analytical perspective, we derived the 
Immirzi parameter, Eq. (\ref{gib}), based on Barrow's statistical model, writing it as a function of Barrow's exponent $\Delta$ 
and the area 
of the punctured surface. In the limit $\Delta \rightarrow 0$, we recovered the usual Immirzi parameter $\gamma = \ln 2 /(\pi \sqrt{3})$. 
Finally, we derived the Immirzi parameter from a modified Kaniadakis entropy, resulting in Eq. (\ref{imike}), which 
represents it as a function of the $\kappa$ parameter and the area of the punctured surface. When $\kappa \rightarrow 0$, we again 
obtained the usual Immirzi parameter $\gamma = \ln 2 /(\pi \sqrt{3})$.
Therefore, by connecting the Immirzi parameter not only to the usual association with the Bekenstein-Hawking entropy law but also to both 
Barrow's framework and a modified Kaniadakis statistics, our approach reinforces the importance of the Landauer principle and reveals new 
perspectives for its application in black hole physics.

\section{Acknowledgments}  
We would like to thank the anonymous Referee for the important and constructive suggestions, which helped us to improve the clarity and scope of 
our manuscript.
Jorge Ananias Neto would like to acknowledge CNPq (Conselho Nacional de Desenvolvimento Cient\'ifico e Tecnol\'ogico), Brazilian scientific support 
federal agency, for partial financial support, CNPq-PQ, Grant number 305984/2023-3.


\begin{thebibliography}{52}

\bibitem{lqg} A. Ashtekar, J. Baez, K. Krasnov, Adv. Theor. Math. Phys. 4 (2000) 1.

\bibitem{lqg2} C. Rovelli and L. Smolin, Nucl. Phys. B 442 (1995) 593.

\bibitem{lqg3} A. Ashtekar and J. Lewandowski, Class. Quantum Gravity 14 (1997) A55.

\bibitem{lqg4} M. Sadiq, Phys. Lett.B 741(2015) 280.

\bibitem{immi} G. Immirzi, Nucl. Phys. B, Proc. Suppl. 57 (1997) 65.

\bibitem{barb} J. Fernando Barbero G.,  Phys. Rev. D 51 (1995) 5507.

\bibitem{dreyer} O. Dreyer, Phys. Rev. Lett. 90(8)  (2003) 081301-1.

\bibitem{landauer} R. Landauer, IBM J. Res. Dev. 5 (1961) 183.

\bibitem{bri} L. Brillouin, J. Appl. Phys. 24 (1953) 1152.

\bibitem{land2} R. Landauer, Phys. Lett. A 217 (1996) 188.

\bibitem{bap} A. Bérut, A. Arakelyan, A. Petrosyan, S. Ciliberto, R. Dillenschneider and E. Lutz,  Nature 483 (2012) 187.

\bibitem{jgb} Y. Jun, M. Gavrilov and J. Bechhoefer,  Phys. Rev. Lett. 113 (2014) 190601.

\bibitem{yxr} L. L. Yan, T. P. Xiong, K. Rehan, F. Zhou, D. F. Liang, L. Chen, J. Q. Zhang, W. L. Yang, Z. H. Ma and M. Feng,
Phys. Rev. Lett. 120 (2018) 210601.

\bibitem{rw} D. Reeb and M. M. Wolf, New J. Phys. 16 (2014) 103011.

\bibitem{ebor} E. Bormashenko, Entropy 21(10) (2019) 918.

\bibitem{ebor2} E. Bormashenko, Entropy 26(5) (2024) 423.

\bibitem{cl} M. Cort\^es and A. Liddle, ``Hawking evaporation and the Landauer Principle", arXiv: 2407.08777v2 [gr-qc].

\bibitem{abreu} E. M. C. Abreu, ``Barrow black hole variable parameter model connected to information theory", 
arXiv: 2402.15922v1 [gr-qc].

\bibitem{otr} Oem Trivedi, ``Universality of Information Thermodynamics and the Efficiency of Information Erasure on the 
Cosmological Apparent Horizon", arXiv: 2407.15231v2 [gr-qc].

\bibitem{bgs} Bijan Bagchi, Aritra Ghosh and Sauvik Sen, Gen. Relativ. Gravit. 56 (2024) 108.

\bibitem{barrow} J. D. Barrow, Phys. Lett. B 808 (2020) 135643.

\bibitem{sari} E. N. Saridakis, Phys. Rev. D 102 (2020) 123525.

\bibitem{sari2} E. N. Saridakis and S. Basilakos, Eur.Phys.J.C 81 (2021) 644.

\bibitem{aab} E. M. C. Abreu, J. Ananias Neto and E. M. Barboza, EPL 130 (2020) 40005.

\bibitem{kj} R. K. Kaul and P. Majumdar, Phys. Rev. Lett. 84 (2000) 5255.

\bibitem{tsallis} C. Tsallis, J. Stat. Phys. 52 (1988) 479.

\bibitem{ww} G. Wilk and  Z. Włodarczyk. Phys. Rev. Lett. 84 (2000) 2770.

\bibitem{tsallis2} C. Tsallis and L. J. L. Cirto, Eur. Phys. J. C 73 (2013) 2487.

\bibitem{Abreu:2020dyu} E.~M.~C.~Abreu and J.~Ananias~Neto, Eur. Phys. J. C 80, no.8, 776 (2020).

\bibitem{bimmi} E. M. C. Abreu and J. Ananias Neto, Phys. Lett. B 807 (2020) 135602.

\bibitem{k1} G. Kaniadakis, Physica A 296 (2001) 405.

\bibitem{k2} G. Kaniadakis, Phys. Rev. E 66 (2002) 056125.

\bibitem{k3} G. Kaniadakis, Phys. Rev. E 72 (2005) 036108.

\bibitem{k4} G. Kaniadakis, Entropy 26 (2024) 406.

\bibitem{ks} G. Kaniadakis and A. M. Scarfone, Physica A 305 (2002) 69.

\bibitem{kqs} G. Kaniadakis, P. Quarati and A. M. Scarfone, Physica A 305 (2002) 76.

\bibitem{aabn} E. M. C. Abreu, J. Ananias Neto, E. M. Barboza and R. C. Nunes, Physica A 441 (2016) 141.

\bibitem{aamp} E. M. C. Abreu, J. Ananias Neto, A. R. Mendes and R. M. de Paula, Chaos, Solitons and Fractals 118 (2019) 307.

\bibitem{lbs} A. Lymperis, S. Basilakos and Emmanuel N. Saridakis, Eur. Phys. J. C 81 (2021) 1037.

\bibitem{ggl} G. G. Luciano, Eur. Phys. J. C 82 (2022) 314.

\bibitem{as} A. Sheykhi, Phys. Lett. B 850 (2024) 138495.

\bibitem{aa} E. M. C. Abreu and J. Ananias Neto, EPL 133 (2021) 49001.

\bibitem{aaacnt} G.~V.~Ambr\'osio, M.~S.~Andrade, P.~R.~F.~Alves, C.~N.~Costa, J.~Ananias~Neto and R.~Thibes,
Braz. J. Phys. 55 (2025) 5, 212.

\bibitem{calabrese} P. Calabrese and J. Cardy,  J. Phys. A: Math. Theor. 42 (2009) 504005.

\bibitem{nishi} T. Nishioka, Rev. Mod. Phys. 90 (2018) 035007.

\bibitem{ci} V. G. Czinner and H. Iguchi, Phys. Lett. B 752 (2016) 306.

\bibitem{ci2} V. G. Czinner and H. Iguchi, Eur. Phys. J. C 85 (2025) 443.

\bibitem{phl} C. Promsiri, E. Hirunsirisawat and W. Liewrian, Phys. Rev. D 102 (2020) 064014.


\end{thebibliography}
\end{document}